\documentclass[10pt]{article}
\usepackage{graphicx,floatflt,amssymb,epsfig} 
\def\be {\begin{equation}}
\def\ee {\end{equation}}
\def\bea {\begin{eqnarray}}
\def\eea {\end{eqnarray}}

\def\slash {\!\!\!/}

\def\del {\partial}

\def\ztwo {{\cal Z}_2}

\def\stwo {\sqrt{2}}

\def\mhsbar {\overline{m_h^2}}

\def\opcit(#1){ {\em op. cit.}, #1}
\def\etal {\em et al.}

\def\issue(#1,#2,#3){{\bf #1}, #2 (#3)} % AIP format

\def\APP(#1,#2,#3){Acta Phys.\ Polon.\ \issue(#1,#2,#3)}
\def\ARNPS(#1,#2,#3){Ann.\ Rev.\ Nucl.\ Part.\ Sci.\ \issue(#1,#2,#3)}
\def\CPC(#1,#2,#3){Comp.\ Phys.\ Comm.\ \issue(#1,#2,#3)}
\def\CIP(#1,#2,#3){Comput.\ Phys.\ \issue(#1,#2,#3)}
\def\EPJC(#1,#2,#3){Eur.\ Phys.\ J.\ C\ \issue(#1,#2,#3)}
\def\EPJD(#1,#2,#3){Eur.\ Phys.\ J. Direct\ C\ \issue(#1,#2,#3)}
\def\IEEETNS(#1,#2,#3){IEEE Trans.\ Nucl.\ Sci.\ \issue(#1,#2,#3)}
\def\IJMP(#1,#2,#3){Int.\ J.\ Mod.\ Phys. \issue(#1,#2,#3)}
\def\JHEP(#1,#2,#3){J.\ High Energy Physics \issue(#1,#2,#3)}
\def\JPG(#1,#2,#3){J.\ Phys.\ G \issue(#1,#2,#3)}
\def\MPL(#1,#2,#3){Mod.\ Phys.\ Lett.\ \issue(#1,#2,#3)}
\def\NP(#1,#2,#3){Nucl.\ Phys.\ \issue(#1,#2,#3)}
\def\NIM(#1,#2,#3){Nucl.\ Instrum.\ Meth.\ \issue(#1,#2,#3)}
\def\PL(#1,#2,#3){Phys.\ Lett.\ \issue(#1,#2,#3)}
\def\PRD(#1,#2,#3){Phys.\ Rev.\ D \issue(#1,#2,#3)}
\def\PRL(#1,#2,#3){Phys.\ Rev.\ Lett.\ \issue(#1,#2,#3)}
\def\SJNP(#1,#2,#3){Sov.\ J. Nucl.\ Phys.\ \issue(#1,#2,#3)}
\def\ZPC(#1,#2,#3){Zeit.\ Phys.\ C \issue(#1,#2,#3)}

\begin{document} 
\begin{flushright} 
CU-PHYSICS/14-2008\\
\end{flushright} 
\vskip 30pt 
 
\begin{center} 
{\Large \bf Universal extra dimension: Violation of Kaluza-Klein parity}\\
\vspace*{1cm} 
\renewcommand{\thefootnote}{\fnsymbol{footnote}} 
{\large {\sf Biplob Bhattacherjee}  } \\ 
\vspace{10pt} 
{\small 
   {\em Department of Physics, University of Calcutta, 92 A.P.C. 
        Road, Kolkata 700009, India}}\\
 E-mail: {\sf bbhattacherjee@gmail.com}
\normalsize 
\end{center} 
 
\begin{abstract} 
The minimal Universal Extra Dimension (mUED) model respects the Kaluza-Klein 
(KK) parity $(-1)^n$, where $n$ is the KK number. However, it is possible to 
have interactions located at only one of the two fixed points of the 
$S_1/\ztwo$ orbifold. Such asymmetric
interactions violate the KK parity. This kills the cold dark matter 
component of UED but also removes the upper bound on the inverse 
compactification radius, and thus non-observation of the KK excitations 
even at the Large Hadron Collider does not necessarily invalidate the model. 
Apart from the decay of the lightest $n=1$ KK
 excitation, this leads to collider signals which are markedly different from 
those in the mUED scenario. The phenomenological consequences of such KK-parity
 violating terms are explored.
 
\vskip 5pt \noindent 
%\texttt{PACS numbers:~ PACS numbers here} \\ 
\texttt{Keywords:~~Universal Extra Dimension, KK Parity violation, Large Hadron
 Collider }
\end{abstract}

\renewcommand{\thesection}{\Roman{section}} 
\setcounter{footnote}{0} 
\renewcommand{\thefootnote}{\arabic{footnote}} 

\section{Introduction}

The Universal Extra Dimension model (UED) is one of the minimal possible 
extensions of the Standard Model (SM) with one or more compactified extra 
dimensions. The minimal model, henceforth called mUED, was proposed by 
Appelquist, Cheng and Dobrescu \cite{acd}, and assumes that all SM fields
can propagate in one compactified extra dimension $y$. To get chiral fermions
at the zero-th level, 
one needs an $S_1/\ztwo$ orbifolding, with two fixed points at $y=0$ and $y=\pi
R$, where $R$ is the compactification radius.  
All extra dimensional models are nonrenormalisable, and can at best be treated
as an effective theory valid upto a cut-off scale $M_s$. Thus, mUED has two 
free parameters, $R$ and $M_s$, with which one can specify the spectrum at
any $n$ level. 
(The masses of the excited scalars also depend on the SM Higgs boson mass 
$M_h$, so strictly speaking, that is also an input parameter.)

One of the interesting feature of the mUED model is the conservation of the KK
number. This comes from the fact that all particles can propagate in the extra
dimension and so the momentum along the fifth dimension must be
conserved. However, presence of two fixed points breaks the translational
symmetry along $y$, so the KK number $n$ is no longer conserved. 
 In principle, there may exist some interactions located only at these fixed
  points. If the interactions are symmetric under the exchange of the fixed 
  points (this is another $\ztwo$ symmetry, but not the $\ztwo$ of 
  $y\leftrightarrow -y$), the conservation of KK number breaks down to the 
  conservation of KK parity, defined as $(-1)^n$, where $n$ is the KK
number\cite{cms1}.
The mUED assumes the conservation of KK parity. KK parity does not
 allow single production of $n=1$ paricles and guarantees the stability of
the lowest lying $n=1$ KK state (LKP). The LKP, for most of the parameter
space, is an excitation of the hypercharge gauge boson $B$ and is an excellent
cold dark matter (CDM) candidate \cite{Feng}. The mass of LKP is approximately
$1/R$ and
hence the overclosure of the universe puts an upper bound on $R^{-1} \leq
800$ GeV \cite{hooper}. This guarantees the production of at least the $n=1$
excited states 
at the Large Hadron Collider (LHC). 

While the spectrum for any $n$ is highly degenerate at the tree-level, the
radiative corrections lift the degeneracy and provide interesting collider
phenomenology \cite{cms1,cms2}. 
There are two types of corrections: the first one, 
which results just from the compactification of the extra dimension, called 
bulk correction, is in general small (zero for fermions) and is constant
for all $n$ levels. The second one, called boundary correction, 
is comparatively large (goes as $\ln(\Lambda^2/q^2)$), and plays the 
major role in 
determining the exact spectrum and possible decay modes. In the mUED model,
it is assumed that the boundary corrections vanish at $q=M_s$, and so
one can identify the cut-off scale $M_s$ with the regularization scale
$\Lambda$.

The collider phenomenology of minimal UED has been investigated in detail
 \cite{lhc-ued,susy-ued,rizzo,bk023}.
A crucial feature of these studies is the existence of the LKP and
hence the missing energy and missing transverse momentum signal at the
colliders, which stems from the conservation of KK parity. However, it is
possible to have fixed-point located interactions that are asymmetric in nature
\cite{acd}. This violates the KK parity, analogous to the R parity violation in
supersymmetry. Some phenomenology of such KK parity violation were also
discussed in \cite{rizzo,Aguila}. The possibility of KK parity violation
with a `partial' universality (by not allowing, or only partially allowing,
 the Higgs boson to flow in the fifth dimension) has also been discussed by
\cite{nandi:fatbrane}. KK-parity violation, in the context of other extra
dimensional models, has also been considered in \cite{Mohapatra}.

The aim of this paper is to investigate the phenomenology of such a KK parity
violating (KKPV) model where a fixed-point located  asymmetric term 
is responsible for KKPV. Unless the KKPV couplings are uninterestingly tiny,
the LKP will decay within the lifetime of the universe and will not be a CDM
candidate anymore. While this removes one of the main motivations of UED, 
this also enlarges the parameter space by removing the upper bound
coming from the CDM density. In other words, we plan to answer this
question: {\em If LHC does not find a KK excitation within the overclosure
bound, which, taking into the possibility of a heavy SM Higgs boson and the
subsequent relaxation of the said bound, is about 1.4 TeV \cite{senami}, 
does it mean
that the UED model is ruled out?} We will show that the answer is negative,
and point out the major signals coming from such a model.

A major spin-off of the model is the possibility to accommodate graviton
states. In the conventional UED model with conservation of KK parity,
gravitons cannot be accommodated for $R^{-1}\leq 800$ GeV \cite{Feng} as
the model would allow an unacceptably large rate of $\gamma_1\to \gamma+
G_1$ (where $G_1$ is the first excited graviton state) which is ruled out
from the cosmological diffuse photon flux. However, if KK parity is violated,
the excited photons would decay to conventional fermion-antifermion pairs and
not to gravitons, as the latter process is suppressed by the Planck mass. 
Thus, there are no dangerous gravitons in the model. 

The plan of this article is as follows. We first describe some of the basic 
features of UED model required for our analysis. Readers familiar with the 
formalism of UED can directly go to Section 3, where the KK parity violation 
is introduced. In Section 4, the nature and various decay modes of the 
erstwhile stable LKP, and also the NLKP (next-to-lightest KK particle)
are discussed. We show the various possible combinations of LKP and 
NLKP depending on the model parameters. In Section 5, we briefly touch upon the
collider signatures of such modes at the LHC. Finally Section 6 summarizes the
results and addresses the possible issues of this work.

\section{The mUED model}
The model has been discussed in great detail in the literature.
In this section we briefly mention some of the interesting features of UED 
required for our analysis.
\begin{itemize}
\item 
The tree level mass spectrum for any level $n$ is almost degenerate. The 
masses of these KK modes are given (at tree level) by $M_n^2 = M_0^2 + 
\left(nR^{-1}\right)^2$, where $M_0$ is the mass of the corresponding 
SM particle. The tree level relation is modified when radiative 
corrections are taken into account\cite{cms1,georgi,pk}. This causes significant
splitting among 
the particle masses of any KK level and has important effects on collider 
phenomenology. The one loop corrected masses are determined by
$R^{-1}$ and $\Lambda$, the cut off scale. In order to determine the 
excited scalar masses the SM Higgs mass ($M_h$) is also required.

\item 
All $n=1$ particles have to be pair produced and ultimately they must cascade
down to the lowest lying $n=1$ particle (LKP), due to the conservation of 
KK parity. However, as the mass splitting among the $n=1$ states is
 generally small (being induced by radiative corrections), 
the final state will be soft leptons or jets associated with missing 
transverse momentum. While the missing energy is large, the missing $p_T$ is
small, because the spectrum is still quasi-degenerate.

\item 
For most of the parameter space ($R^{-1} \leq 800$ GeV) the $n=1$ photon, which
is almost the excitation of the hypercharge gauge boson $B$ due to the smallness
of the `Weinberg angle' for the $n>0$ levels, is the LKP. (If gravitons are
included, they become the LKP and the model runs into trouble with the diffuse
photon flux coming from $B_1 \to \gamma+ G_1$, so we exclude gravitons).
For higher values of $R^{-1}$ and for high $M_h$, $H^\pm_1$, the charged Higgs
boson excitation, becomes the LKP. This possibility cannot be encouraged if LKP
is stable and a prospective CDM candidate. 

\item  
The corrections for KK modes with electroweak interactions are generally small.
Singlet or doublet leptons lie just above the $B_1$. Masses of the three
excited scalars $h_1$, $A^{0}_1$, and $H^{\pm}_1$ are also very close to the 
excited lepton masses (see, {\em e.g.}, \cite{cms1} for a benchmark spectrum).
However, the scalar masses depend on the SM Higgs mass $M_h$;
if we keep $R^{-1}$ and $\Lambda$ fixed, for larger $M_h$, $H_1^\pm$ 
and $A_1^0$ masses go down and $h_1$ becomes more massive. 
The charged scalar mass can even go below the $B_1$ mass \cite{Feng}.

\item 
The mass equation for the excited scalars may contain a universal 
boundary-located soft term $\mhsbar$ \cite{cms1}. This term is a free parameter of the 
theory and is taken to be zero in the mUED model. 
For fixed $M_h$, all excited scalar masses increase with increasing $\mhsbar$. 
However, large negative values of $\mhsbar$ again drives the $H^\pm_1$ to 
be the LKP (this gives a lower limit on $\mhsbar$ as a function of $R$, 
$\Lambda$ and $M_h$) \cite{bbscalar}. 

\item 
The $n=1$ fermions are vectorial and can be $\ztwo$-even 
(left doublet and right singlet) or $\ztwo$-odd (left singlet and right 
doublet). These states are not exactly the mass
eigenstates. In the doublet-singlet basis, we get a non-diagonal mass matrix,
whose off-diagonal entries are the zeroth level mass $M_0$. So the mass 
matrix is almost diagonal for all fermions, except for the third generation 
quarks, in particular for the top. After diagonalisation and a chiral
 rotation, one gets the proper mass eigenstates.

\end{itemize}

\section{KK parity violation}

In this section we study the effect of localised kinetic operators on the
boundary. The possibility of such terms has been mentioned in
\cite{acd,rizzo,Aguila}.
Let us consider, as an illustrative example, the simplest possible one, 
{\em i.e.}, the fermion kinetic term, which, located at  $y=y_0$, looks like 
\begin{equation}
L_{f}= {\lambda \over{2 M_s }}\int{[i\bar{\psi}{\Gamma^{\alpha}} { D_{\alpha}}
{\psi} } -i
({D_\alpha ^{\dagger} } \bar{\psi}) \Gamma^{\alpha} \psi]\delta(y-y_0)~dy\,,
 \label{int}
\end{equation}
 where $\psi(x^\mu,y)$ is any five dimensional fermionic field and 
$\lambda$ is the coupling constant. The term is suppressed by the cutoff scale 
$M_s$, which may be identified with $\Lambda$.
Such a term contributes to the kinetic term of the KK fermion and hence
changes its mass. To determine the spectrum, we have made several 
simplifying assumptions, without seriously compromising with the phenomenology. 

\begin{itemize}
\item 
The first assumption is to
place such a term only at $y=0$ without losing any generality as the points
$y=0$ and $y=\pi R$ have already been chosen as the fixed points for the 
orbifold and the decomposition into even and odd modes are performed 
accordingly. For this case, only even modes will mix with each other. 
%If we take other point except the fixed point, all even and odd states would be mixed.

\item 
To concentrate on the LHC-related phenomenology, let us consider the mixing
between $n=0$ and $n=1$ states only. In principle, all $n$ states can mix
with each other, but the admixture of the higher states in the low-lying 
physical states are suppressed by their masses.

\item 
We take the KK parity violating effects, parametrised by the dimensionless
coupling
\begin{equation}
h = \frac{\lambda}{2\pi M_s R}\,,
\end{equation}
to be small (it should be perturbative for any meaningful calculation). In fact,
we would take $h$ to be so small, ${\cal O}(10^{-2})$, that effects on the
spectrum that depends quadratically on $h$ can be neglected. 
In this limit, it is enough to compute the tree-level corrections to the
spectrum and neglect the loop effects.

\item 
We take $\lambda$ and hence $h$ to be the same for all fermionic flavours.
This is in conformity with lepton universality and suppression of tree-level
FCNCs. On the other hand, non-uniformity of $\lambda$ may be constrained from 
such low-energy observables, in analogy with the R-parity violating couplings of
supersymmetry. 

\item 
If $h$ is small, we can take the standard KK expansion of the fields 
as defined in \cite{acd}. For large values of $h$, the expansion is most
definitely not valid, but if we neglect terms ${\cal O}(h^2)$ and higher,
one can use the standard perturbation theory with the $h=0$ limit as the
unperturbed basis.

\end{itemize}

We now integrate over $y$ and get the usual 4-d Lagrangian. The kinetic part of
 the 4-d Lagrangian in the ($n=0$, $n=1$ doublet, $n=1$ singlet)
basis is given by 
\be
{i\over 2} \pmatrix{ \bar{\psi_L^{(0)}} & \bar{\psi_L^{(1)}} &
\bar{S_L^{(1)}}}K_L
i\gamma^\mu\del_\mu\pmatrix{\psi_L^{(0)} \cr \psi_L^{(1)} \cr S_L^{(1)}}+
{i\over 2} \pmatrix{ \bar{\psi_R^{(0)}} & \bar{\psi_R^{(1)}} &
\bar{S_R^{(1)}}}K_R 
i\gamma^\mu\del_\mu\pmatrix{\psi_R^{(0)} \cr \psi_R^{(1)} \cr S_R^{(1)}}\, +
h.c ,
\ee
where 
\be
K_L=
\pmatrix{1/2+ h & \stwo h & 0\cr
\stwo h &  1/2+2h & 0\cr
0 & 0 & 1/2 }\,; \ \ 
K_R=
\pmatrix{1/2+ h &0 & \stwo h \cr
0 & 1/2  & 0 \cr
\stwo h & 0 & 1/2+2h }\,.
\ee

Note that the odd fields $S^{(1)}_L$ and $\psi^{(1)}_R$ do not mix with 
the $n=0$ fields.
Here 
%we consider only mixing between zeroth mode and the first mode with these
%corrections. 
$K_L$ and $K_R$ are two symmetric matrices and they are
diagonalised by two orthogonal matrices $E_L$ and $E_R$. The eigenvalues of
$E_L$ and $E_R$ are $1/2 , 1/2 $ and $(1+6h)/2$. We now rescale the kinetic
terms by two normalisation matrices $N_L$ and $N_R$. After diagonalisation and
rescaling, the kinetic terms take their canonical forms. In this intermediate
basis, the kinetic terms are diagonal but the mass matrix is not; rather, it
is of the form
\be
M^{'} = N_L^{-1} E_L^{T} M E_R N_R^{-1}\,,
\ee
where $M$ is the mass matrix in the KK basis:
\be
M=
\pmatrix{M_0 &h\stwo/R  & 0\cr
0 & 2 h / R + 1/R +\Delta_D  & M_0 \cr
h\stwo/R & M_0 & -2 h / R - 1/R - \Delta_S}
\ee 
where $M_0$ is the $n=0$, {\em i.e.}, SM fermion mass, and $\Delta_D$ and 
$\Delta_S$ are the radiative corrections on $n=1$ doublet and singlet fermions
respectively. Their expressions can be found in {\cite{cms1}} and do not 
change in the limit of small $h$.
$M^{'}$ is neither diagonal nor symmetric but can be diagonalised by a 
bi-unitary transformation of the form
$M_D=V^{\dagger} M^{'} U $. The unitary matrices $U$ and $V$ can be obtained
by diagonalising $M^{'}M^{'\dagger}$ and $M^{'\dagger}M^{'}$ respectively. 
We can choose $U$ and $V$ in such a way that all elements of $M^{'}$  are 
positive.

The transformation equations which connect original KK basis to the mass basis are given below
\be
\pmatrix{\psi_L^{(0)} \cr \psi_L^{(1)} \cr S_L^{(1)}}= N_R^{-1} U \pmatrix
{\phi_L^{(1)} \cr \phi_L^{(2)} \cr \phi_L^{(3)}}\,; \ \ 
\pmatrix{\psi_R^{(0)} \cr \psi_R^{(1)} \cr S_R^{(1)}}= N_L^{-1} V\pmatrix
{\phi_R^{(1)} \cr \phi_R^{(2)} \cr \phi_R^{(3)}}\,.
\ee
The transformation is not unitary in nature as the normalisation matrices 
are themselves non-unitary.

Once we obtain the physical states, the Feynman rules can be computed from the
Lagrangian and the mixing matrix. However, for small $h$ so that ${\cal O}(h^2)$
terms can be neglected, the rules are particularly simple, {\em e.g.},
\begin{eqnarray}
\overline{e_1^{(1)}} e^{(0)} \gamma^{(0)} &\Longrightarrow& ieh\sqrt{2} \gamma_\mu
(1+\gamma_5)\,,\nonumber\\
\overline{e_2^{(1)}} e^{(0)} \gamma^{(0)} &\Longrightarrow& ieh\sqrt{2} \gamma_\mu
(1-\gamma_5)\,,\nonumber\\
\overline{e_1^{(1)}} e^{(0)} Z^{(0)} &\Longrightarrow& -\frac{iehs_W\sqrt{2}}{c_W}
 \gamma_\mu (1+\gamma_5)\,,\nonumber\\
\overline{e_2^{(1)}} e^{(0)} Z^{(0)} &\Longrightarrow& \frac{ieh\sqrt{2}}{2s_Wc_W}
 (1- 2 s_W^{2})\gamma_\mu (1-\gamma_5) \,,\nonumber\\
\overline{e_2^{(1)}} \nu^{(0)} W^{(0)} &\Longrightarrow& -\frac{ieh}{s_W}
 \gamma_\mu (1-\gamma_5)\,,\nonumber\\
\overline{e^{(0)}} e^{(0)} B^{(1)} &\Longrightarrow& \frac{ieh\sqrt{2}}{2c_W}\left[
 \gamma_\mu (1-\gamma_5) + 2\gamma_\mu(1+\gamma_5)\right]\,,\nonumber\\
   \label{feynmanrules}
\end{eqnarray}
where the subscripts refer to the dominant admixture at the $n=1$ level, 
and the superscripts to the KK numbers themselves. 
We have not shown the KKPV contributions to the KK-number conserving vertices
({\em e.g.}, an $n=0$ gauge boson coupling to two $n=1$ fermions), as the standard
gauge coupling is overwhelmingly dominant. By the same argument, KKPV decays of
an $n=1$ $W$, $Z$, or gluon to two $n=0$ fermions have not been shown.

\subsection{Spectrum}

With $\lambda\sim 1$, $M_sR\sim 10$, $h_{max}\sim 0.02$. We study the spectrum
by varying $h$ between $-0.02$ and $0.02$. Note that while this range does
not depend on the precise value of $R^{-1}$, the low-energy constraints
should depend upon $hR$ and hence with large values of $R^{-1}$, most of
these constraints could be successfully avoided. For such small values of $h$,
the branching fractions of KK allowed channels will be hardly affected 
(except for a few cases that we will show later), since
they are mostly driven by gauge or large Yukawa couplings. However, the LKP
will decay, and decay promptly within the detector unless $h$ is very tiny.

The mixing angles between the KK basis and the mass basis depend upon $R^{-1}$
and $h$, and also indirectly on $\Lambda$ through $\Delta_D$ and $\Delta_S$.
They also explicitly depend on $M_0$, numerically important only for the top
quark and the otherwise closely spaced levels.  

\begin{figure}[ht]
\centering
\vskip 19pt
\hspace*{0.2cm} 
\psfig{figure=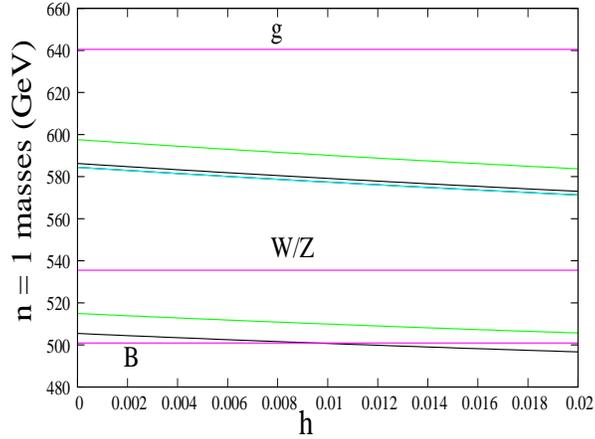,width=7.8cm,height=5.8cm,angle=0}
\caption{Spectrum for the $n=1$ level. From top to bottom, the sloping lines
are for $u_2/d_2$, $u_1$, $d_1$, $l_2/\nu_2$, and $l_1$ respectively. For
the notation, see text.}
\label{fig:spectrum}
\end{figure}

The spectrum for $R^{-1}=500$ GeV is shown in fig.\ \ref{fig:spectrum}. 
The horizontal lines, from top to bottom, stand for excited gluon $g_1$,
$W_1/Z_1$ (they are almost degenerate), and $B_1$ respectvely.
As expected, the excited gauge boson masses are not affected. 
There are two quark states for each flavour; they are the linear combinations of
the singlet and the doublet fields. We shall call them $q_1$ (dominantly 
singlet) and $q_2$ (dominantly doublet). 
{\em{For all fermions, the subscript refers to the SU(2) gauge quantum 
number, not the KK number, but this should not create any confusion as
we are interested in the phenomenology of only the $n=1$ level.}}
The falling lines, from top to bottom, correspond to $u_2/d_2$, $u_1$, 
$d_1$ quarks and $l_2/\nu_2$, $l_1$ leptons respectively.
Note that the fermion masses increase for
negative $h$ and decrease for positive $h$. For the quarks, the change is about
4\%, but there is no level crossing, and hence the decay patterns remain 
identical. (The channel $q_1 \to q_0 V_0$ or $g_1 \to q_0 q_0$ 
opens up, where $V_0$ is a SM gauge
boson, but the coupling is suppressed and the branching ratios are only
minutely modified.) Interesting thing happens for leptons, as
their masses are close to the $B_1$ mass, and level crossing may take place.
For example, for $R^{-1}=500$ GeV and $h\approx 0.01$, the dominantly singlet
lepton $l_1$ can become the LKP and its mUED decay channel to $n=0$ lepton
and $B_1$ closes. The only possible channel for $l_1$ to decay is the KK parity
violating one, to a lepton and an $n=0$ electroweak gauge boson. Similarly,
for $h < 0$ the leptons may go above the $n=1$ scalars (whose masses do not
depend on $h$) and the scalar decay channels undergo a fundamental change,
from two-body $\tau$ modes to three-body $f\bar{f}B_1$.

%%%%%%%%%%%%%

\section{LKP and NLKP: Phase diagram and decay patterns}

If the LKP is no longer the dark matter candidate, it need not be a neutral one
any more. Also, depending on the parameters of the model, namely, $R^{-1}$,
$M_s=\Lambda$, $M_h$, and $h$, there are various possible NLKPs (we take,
for simplicity, $\mhsbar=0$). The possibilities include $l_1$ and $l_2$
$B_1$, $H^\pm_1$, and $A^0_1$ (the neutral CP-odd $n=1$ scalar).

\begin{figure}[htbp]
\vskip 19pt
\hspace*{0.4cm} 
\psfig{figure=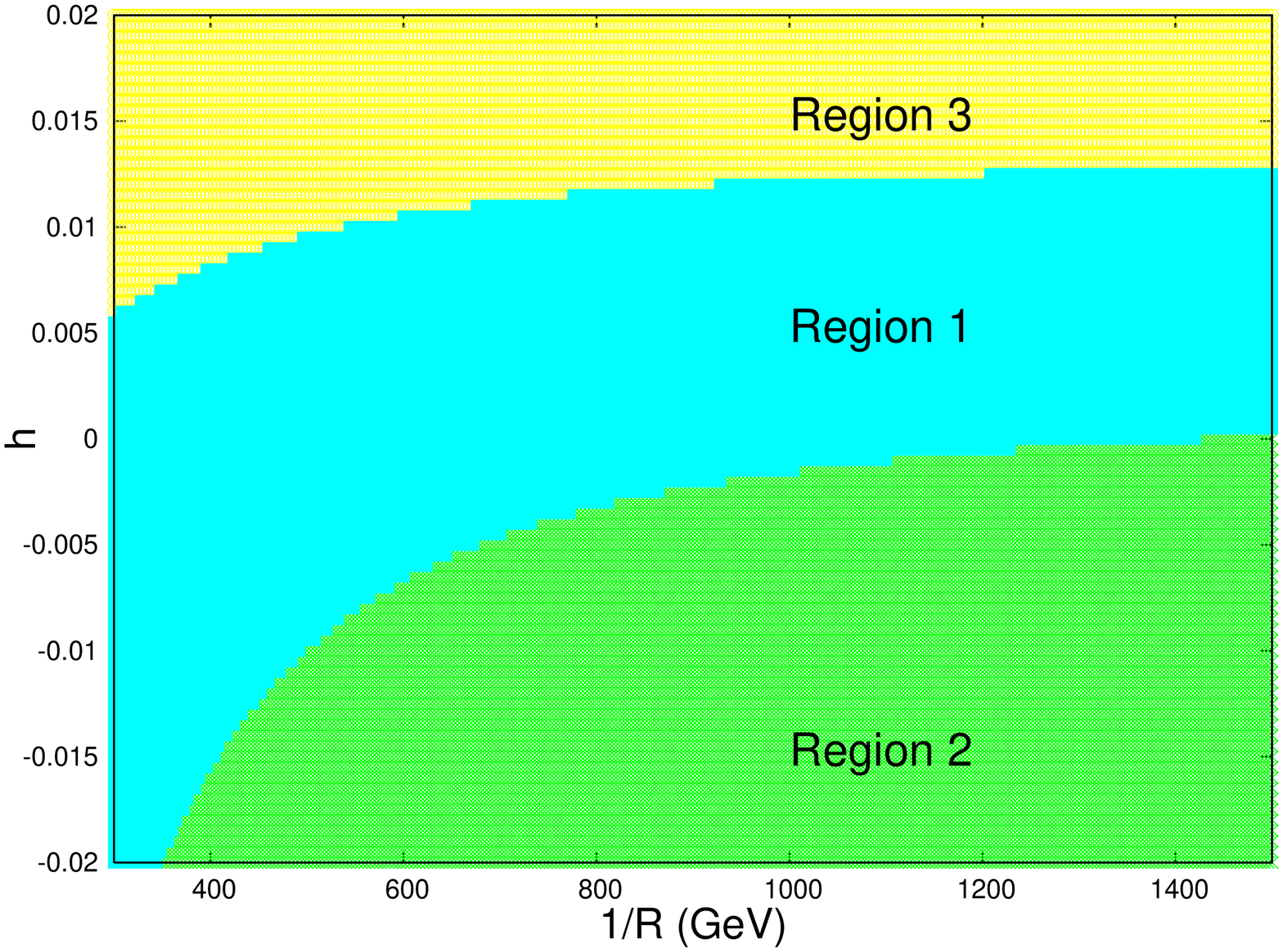,width=6.3cm,height=7.5cm,angle=-90}
\hspace*{0.4cm} 
\psfig{figure= 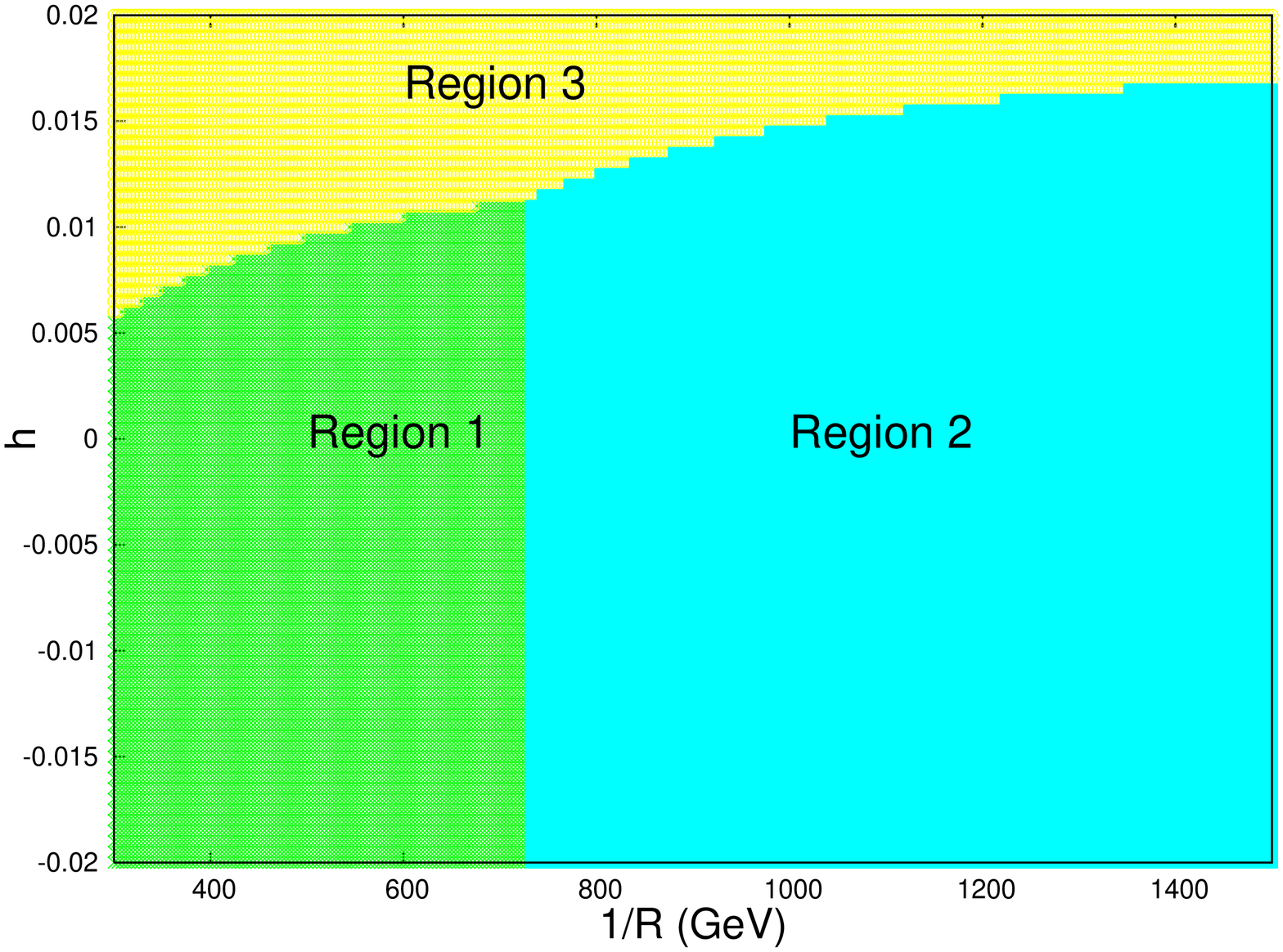,width=6.3cm,height=7.5cm,angle=-90}
\hspace*{.2cm} 
%\psfig{figure=h_200_mh_-10000_bak.eps,width=4.8cm,height=5.8cm,angle=-90}
%\hspace*{-0.2cm} 
\caption{LKP-NLKP phase diagram for $M_h=120$ GeV (left panel) and $M_h=250$
GeV (right panel). For the explanation of the various regions, see text.}
\label{fig:phasedia}
\end{figure}

The LKP-NLKP phase diagrams are shown in fig. \ref{fig:phasedia}, drawn for
$M_s R = 20$, for two 
distinct cases: a light SM Higgs boson ($M_h=120$ GeV) and a heavy SM Higgs
boson ($M_h=250$ GeV). The salient features are as follows:

\noindent{\bf{Case 1}} ($M_h=120$ GeV): 
In this case there are only two possible LKP candidates: $B_1$ (regions 1 and 3)
and $l_1$ (region 2). While the $h=0$ limit corresponds to the mUED ($l_1$
NLKP, region 1), the
transition to $l_1$ LKP can be understood from fig.\ \ref{fig:spectrum}. 
For sufficiently negative values of $h$, $l_1$ goes above $H^\pm_1$, which
then becomes NLKP (region 3). 
The parameter space does not allow $H^{\pm}$ as LKP.
 
\noindent{\bf{Case 2}} ($M_h=250$ GeV): 
For large $M_h$, the situation becomes more complicated. The large quartic
self-coupling drives the $H^\pm_1$ mass down and in region 2, this becomes the
LKP, while it is the usual $B_1$ LKP phase in region 1. In region 1, depending
on the values of the model parameters, either $l_1$ or $H^\pm_1$ can be the 
NLKP, while in region 2, either $B_1$, $A^0_1$, or $l_1$ is the NLKP. Region 3
is the $l_1$ LKP region, with $B_1$ or $H^\pm_1$ as the NLKP. 
The phase diagram can be more complicated with excited gravitons or right-handed
neutrinos, which we have not considered here.

\subsection{Decay of LKP and NLKP}
As was mentioned above, there are three possibilities for the LKP: $B_1$, 
$H^\pm_1$ and $l_1$, depending on the parameter space. The KK parity violating 
interactions allow couplings of $n = 1$ states with two SM particles. There is 
obviously no KK conserving decay modes for LKP.

{\bf {Case 1}} ($B_1$ LKP):  
$B_1$ can decay to two SM fermions, $q\bar{q}$, $l\bar{l}$ or $\nu\bar{\nu}$.
The couplings are proportional to the corresponding hypercharges; the quark
channels are also enhanced by the colour factor. The phase space suppression
is minimal unless $B_1$ is just above the $t\bar{t}$ threshold. 
The table shows the $B_1$ branching fractions for $R^{-1}=500$ GeV (there is
no KK conserving decay, so the widths do not depend on the value of $h$).

\begin{table}[h]
\begin{center}
\begin{tabular}{||c|c|c||}
\hline
 Br($B_1 \rightarrow q \bar{q}$)& 
Br($B_1 \rightarrow e^{\pm}, \mu^{\pm},\tau^{\pm}$) &
 Br($B_1 \rightarrow \nu_e,\nu_\mu, \nu_\tau$)  \\
 ( \% ) & ( \% ) & ( \%) \\
\hline
 53.3     & 38.9    &  7.8 \\
\hline
\end{tabular}
\caption{Branching fractions of $B_1$.}
\end{center}
\end{table}

The hadron channels are difficult to identify at the LHC, except maybe the
$t\bar{t}$ channel. A better option is to look for the dilepton channel 
whose invariant mass peaks at $M_{B_1}$. If $B_1$ is the NLKP then it almost
always decays to $B_1\to l_1 l_0$, followed by the KK parity violating decays
of $l_1$ to $l$ plus $\gamma$ or $Z$ (the $l_0$, being mostly right-chiral 
as it is produced in a vector interaction in association with $l_1$, has a very
small branching ratio to $\nu + W$).

\begin{figure}[htbp]
\centering

\vskip -9pt
\hspace*{0.2cm} 
\psfig{figure=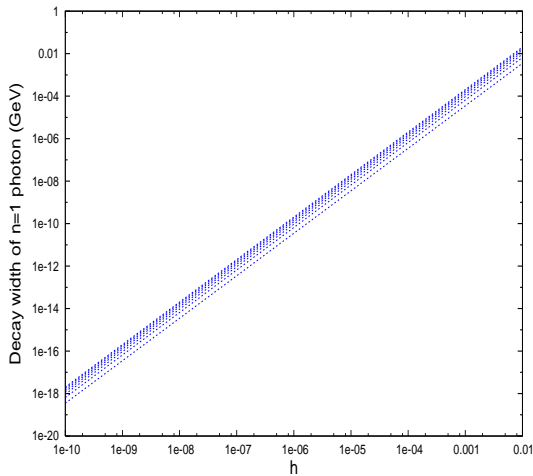,width=6.4cm,height=7.2cm,angle=-90}
\caption{The decay width of $B_1$ as a function of $h$ for different values of
 $R^{-1}$, from 1 TeV (top of the band) to 300 GeV (bottom).}
\label{fig:decayB1}
\end{figure}

There is an interesting possibility. If $h$ is sufficiently small, lifetime 
of $B_1$ may be long enough, so that it can decay outside the detector. Such 
situation can mimic the standard mUED scenario where LKP is stable. One has to 
calculate the lifetime
of $B_1$ as function of $h$ for different values of $R^{-1}$. 
In figure \ref{fig:decayB1} we plot the
two body KK parity violating decay width of $B_1$ as a function of $h$ with 
$R^{-1}$ as a free parameter.
The band corresponds to the variation of $R^{-1}$: the top line for 
$R^{-1} =1$ TeV and the bottom line for $R^{-1} =300$ GeV. $h$ is varied
between $10^{-10}$ to $0.01$. It appears that if $h\sim 10^{-6}$, $B_1$ decay
will lead to a secondary vertex, while if $h\sim 10^{-8}$, the path length is
of the order of a metre and it can decay outside the detector. In this case,
the signals will be identical to that of mUED but the LKP may be considerably 
heavier than the overclosure limit of $\sim 1$ TeV.

%\vskip -150pt

{\bf {Case 2}} ($l_1$ LKP):  
At the $n=1$ levels, there are two leptons, $l_1$ and $l_2$. For small values of
$h$, $l_1$ is dominantly singlet and $l_2$ is dominantly doublet. 
Thus, $l_1$ decays almost entirely to $l+\gamma$ or $l+Z$; the $\nu+W$ channel
will be negligible due to the chiral nature of $l_1$. 

If $l_1$ is the NLKP, it will decay with almost 100\% branching ratio to 
$l+B_1$ if $h$ is small and $l_1$-$B_1$ splitting is at least 1 GeV.
If $h$ is large, the KK parity violating decay modes will start competing with
the KK conserving ones, since the latter becomes more phase space suppressed
with increasing $h$, while the former channel gets enhanced.  
Here we show the KK parity  conserving and KK parity violating branching ratios
 of the singlet lepton for three benchmark values of $h$. In the first case 
$l_1$ is  above the $B_1$ so that KK
conserving decay width is not
suppressed. In the third case $ M_{B_1 } >  M_{l_1} $ so that all KK
conserving channels are closed. In this case it can decay dominantly to SM $Z$
boson. The second case corresponds to the minimal UED where
$h=0$.

\begin{table}[h]
\begin{center}
\begin{tabular}{||c|c|c|c|c||}
\hline
$R^{-1}$ & $h$ & Br($l_1 \rightarrow B_1$)& Br($l_1 \rightarrow \gamma)$ & Br($l_1 \rightarrow Z)$  \\
\hline
500&-0.01 &74.9 &19.6 & 5.5 \\
\hline
500&0.0  &100 &0 & 0 \\
\hline
500 &0.01 &0 &77.9 &22.1 \\
\hline
\end{tabular}
\caption{Branching fractions of $l_1$.}
\end{center}
\end{table}

We also show the decay channels and branching fractions of $l_2$, which can at 
most be an NLKP candidate (it can never go below $l_1$ for small and uniform
values of $h$). Note that the KK parity violating channels may become
important for large $h$. However, neither $l_1$ nor $l_2$ can decay outside the
detector; this happens only for very tiny values of $h$ where $B_1$ is the
LKP.

\begin{table}[h]
\begin{center}
% use packages: array
\begin{tabular}{||c|c|c|c|c|c||}
\hline
$R^{-1}$ & h & Br($l_2 \rightarrow B_1$)& Br($l_2 \rightarrow \gamma)$ & Br($l_2 \rightarrow Z) $ & Br($l_2 \rightarrow W$) \\
\hline
500&-0.01 & 50.6     & 14.3     & 5.5 & 29.6 \\
\hline
500 &0.0 &100 &0 &0 &0 \\
\hline
500& 0.01   &17.9      & 23.8     & 9.1  & 49.1 \\
\hline
\end{tabular}
\caption{Branching fractions of $l_2$.}
\end{center}
\end{table}

{\bf {Case 3}} ($H^\pm_1$ LKP):  
The third possibility, which only occurs for heavy SM Higgs,
is the charged Higgs LKP.
The KK parity violating decays of $H^\pm_1$ and other excited scalars occur 
through the admixture of the $n=1$ state with the physical lowest-lying
state, as there is no such parity violating term for the Yukawa sector
to start with.
$H^\pm_1$ can also decay to $f\bar{f}V_0$, where $V_0$ is a SM gauge boson.
This proceeds through the virtual $n=1$ gauge boson state. The exact branching
fractions depend on $h$; for very small $h$, $H^\pm_1$ can decay outside the
detector and one observes the thick charged track, something reminiscent of
a long-lived chargino.

%%%%%%%%%%%%%%%%%%

\section{Collider Phenomenology}

\begin{figure}[htpb]
\hspace*{-0.5cm}
%\centerline{
\rotatebox{-0}{ \epsfxsize= 8.0 cm\epsfysize=8.0cm \epsfbox{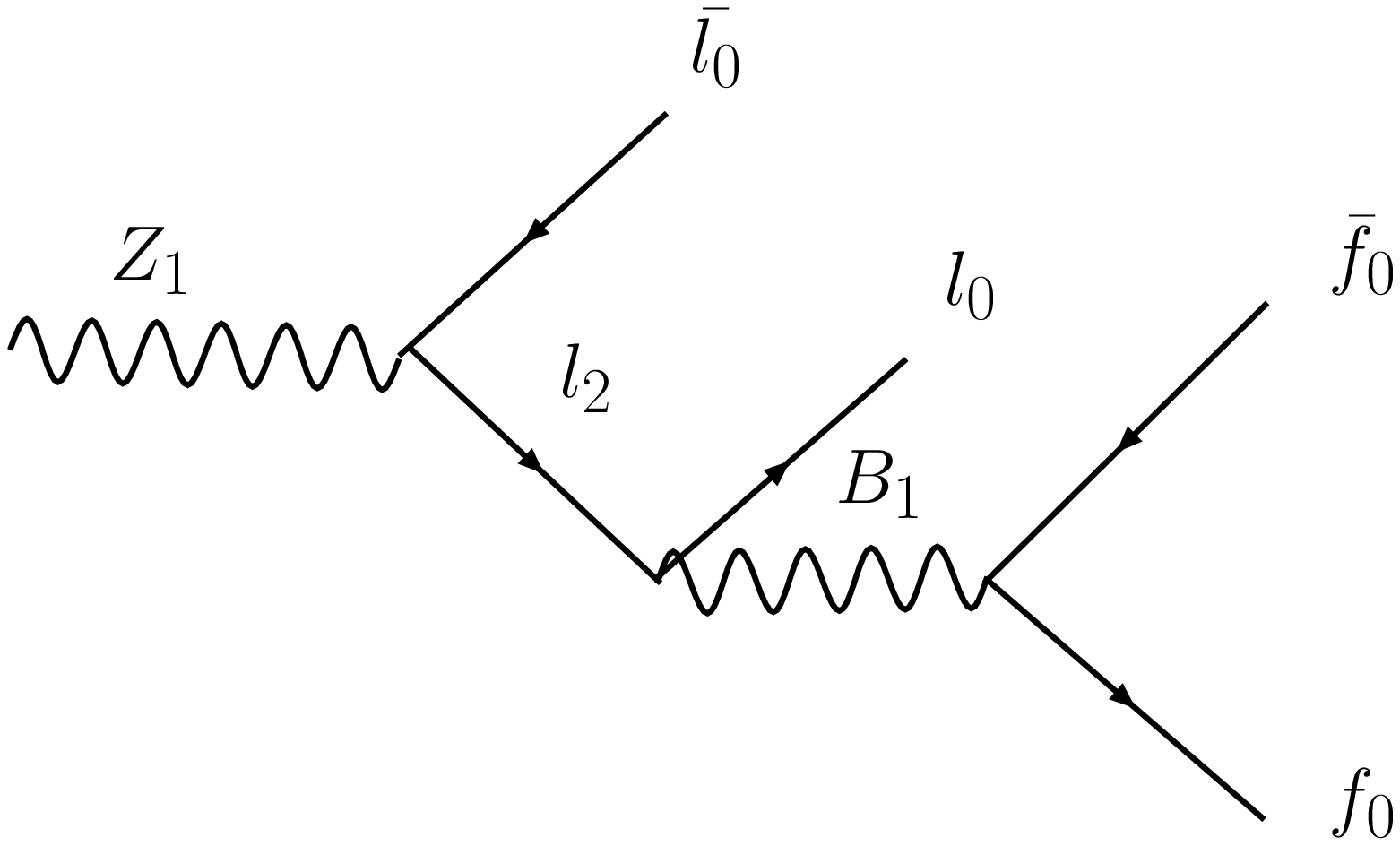} }
\hspace*{-1.5cm}
\rotatebox{-0}{ \epsfxsize= 8.0 cm\epsfysize=8.0cm \epsfbox{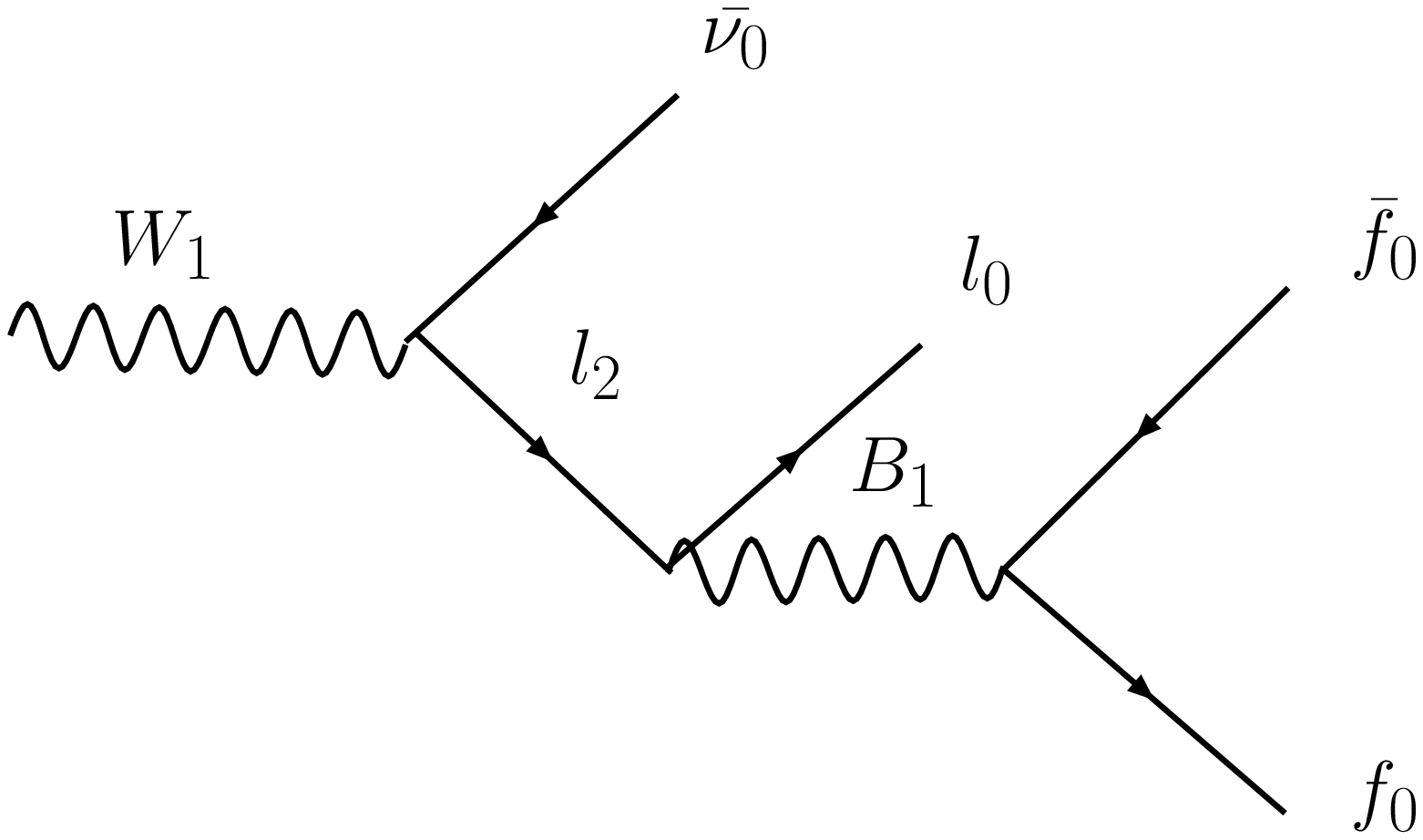} }
\vskip -140pt
\hspace*{.5cm}
\rotatebox{-0}{ \epsfxsize= 6.0 cm\epsfysize=6.0cm \epsfbox{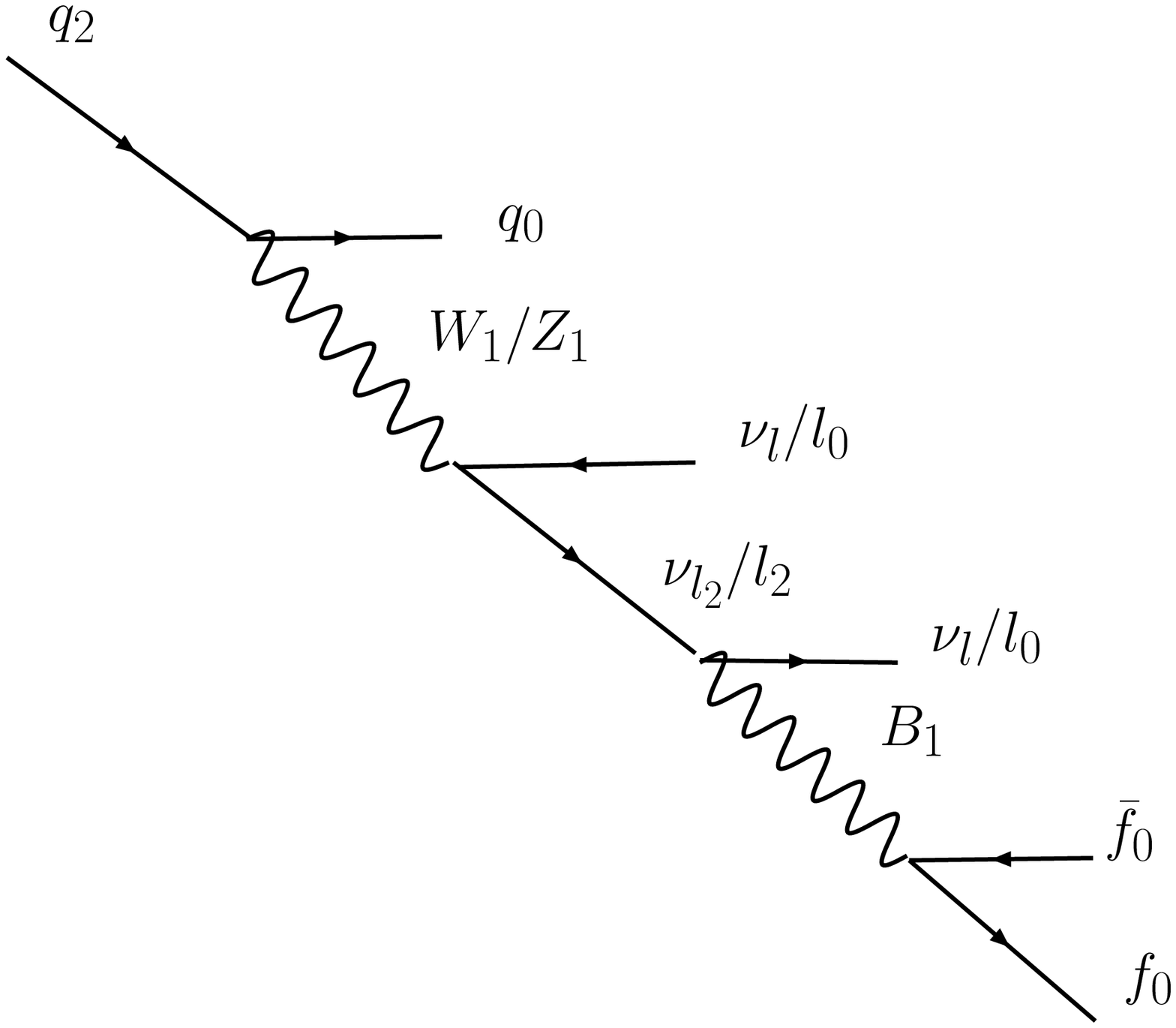} }
\vspace*{-.5cm}
%\vskip -20pt
%\vskip 5pt
\rotatebox{-10}{ \epsfxsize= 8.0 cm\epsfysize=6.0cm \epsfbox{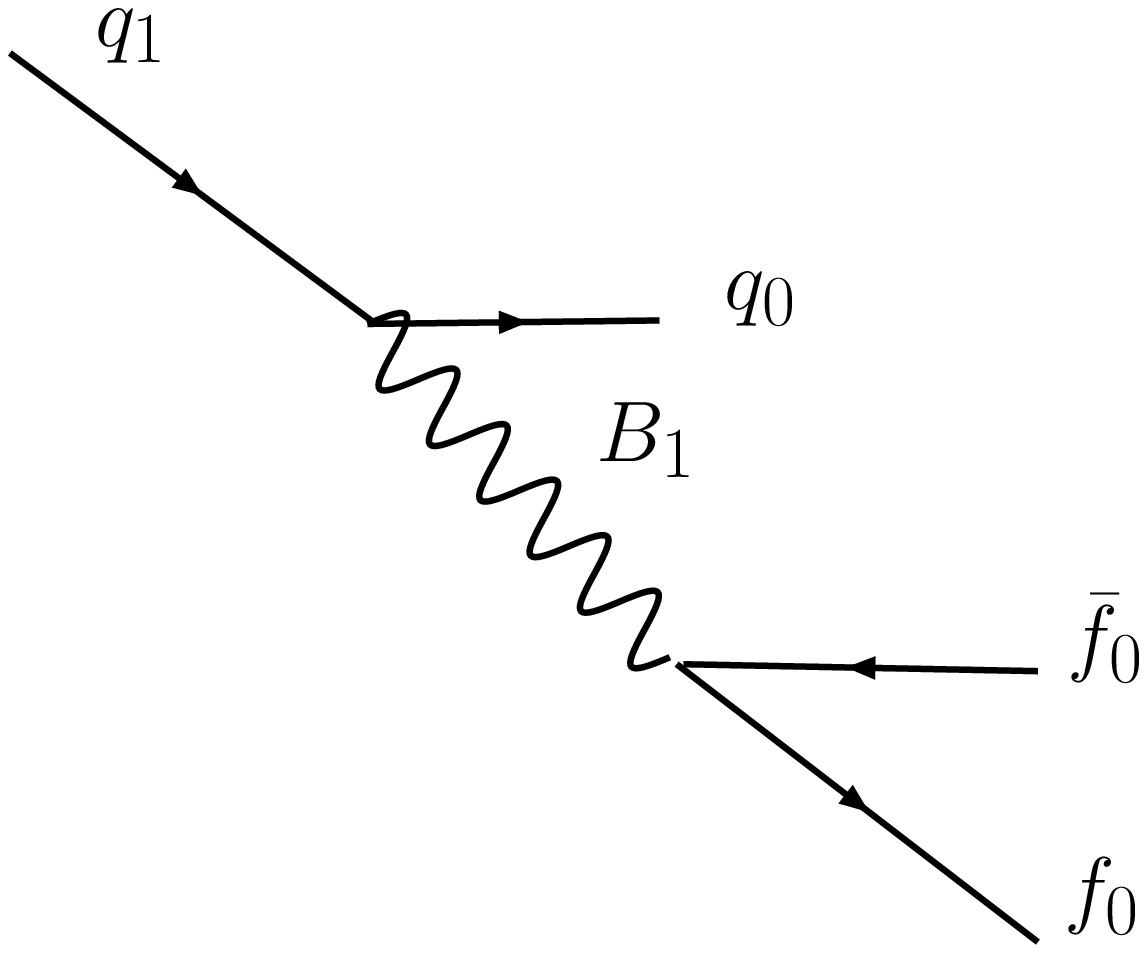} }
%}
\vskip -70pt
\caption{Feynman diagrams of $q_1$, $\tilde{q_1}$, $W_1$ and $Z_1$ decay}
\label{fig:dia}
\end{figure}

The characteristic collider signal of the mUED model is SM 
 particles with low  transverse energy and a huge amount of missing energy,
 which is very similar to the R-parity conserving SUSY models. Missing
 energy comes from the stable neutral LKP, $B_1$, which
 does not interact with the detector. But when KK parity is broken we 
 lose the missing energy part of the signals because LKP is no longer stable.
 The only source of missing energy in this case is the SM neutrino which may
  come from the decay of $B_1, W_1$ and $ Z_1 $ (see fig.\ \ref{fig:dia}).
  In our analysis we consider only those cases where $h$
 is so small that single production of KK excitations 
is not allowed at colliders,
 but the strength is  sufficient  to allow  KK particles to decay within the
 detector. In other words, we do not consider  any
single production but study the effect of KK parity violation at the
last stage of the cascade where LKP is produced. The model follows the mUED
allowed productions and decays except the LKP decay.
Also for small value of $h$, the mass spectum remains almost unchanged.

We are now in a position to discuss qualitatively the experimental signatures 
of such a model.
At the LHC, KK excitations can be produced  mainly through strong interaction.
The dominant processes are the pair production of $n=1$ colored objects: 
$$ 
p p \longrightarrow g_{n=1} q_1/q_2\,,\qquad g_{n=1} g_{n=1}\,, \qquad q_1/q_2 
q_1/q_2\,,
$$
where $g_{n=1}$ is the $n=1$ gluon and  $q_1/q_2$ are the dominantly singlet or
doublet $n=1$ quark states.
The production processes of electroweak strength are
$$
p p \longrightarrow W W\,,\ \ W Z\,, \ \ 
Z Z\,, \ \  {l_1} \bar{l}_1\,, \ \ 
{l_2} \bar{l}_2\,,\ \ {\nu_2} \bar{\nu}_2\,,
$$
where all gauge bosons are the $n=1$ states.
The cascade decays of the produced excited particles result in final state
 with two $B_1$. In the mUED case, $B_1$ is stable and thus escapes the 
detector. The singlet quark can decay only to $B_1$, whereas doublet quarks can 
decay mostly to $W_1$ or $Z_1$. The hadronic
 decay modes of $W_1$ and $Z_1$ are closed, and they decay universally to all 
lepton flavours. 
The leptons finally decay to $B_1$. Thus, the final state signature of 
mUED is $n$ jets + $m$
 leptons + missing $E_T$. 

When KK parity is broken, $B_1$ can decay two SM fermions.
Some of the possible decay chains are shown in figure \ref{fig:dia}.
We get a huge number of different final states depending on the deacy pattern 
of $B_1$.
 The decay of the LKP will increase the particle multiplicity in the final
state and we expect an excess in the SM particles. This
 is  similar to the R-parity violating SUSY scenarios. The decay pattern of 
$B_1$ shows that it can decay invisibly (although the branching is small 
$\sim 8\%$). 
This gives rise to missing energy in the final state. The SM neutrinos may
also come from $W_1$ or $Z_1$ decay.

%%%%%%%%%%%%%%%%
%%%%%%%%%%%%%%%%%%%%

 In the mUED model the spectrum is highly degenerate even after radiative
 correction. So the missing $E_T$ is not large (the distribution has a peak 
at about 100 to 150 GeV).
 The signals with no missing energy are not easy to detect, because of the
huge SM backgrounds. We will
 get 4 lepton or 2 lepton plus missing energy or 2 leptons plus 2 quarks or two quarks plus missing
 energy, or 2 lepton plus 2 quarks from $Z_1$ decay.
$Z_1$ can also decay invisibly, although the branching ratio is small. 
The $W_1$ decays are very similar to the $Z_1$ decay. The final state coming 
from $g_1 g_1$ must contain $N \geq$ 4 jets. It may contain a large number of 
leptons ($\leq$ 8) too, without or with missing energy. If $B_1$ decays to $t 
\bar{t}$ the final states will turn out to be very complicated. If we consider 
the production processes like $g_1 q_1$ or $q_1 q_1$, final state will again be
 multijet ($\geq$ 3 or 2 ) and multilepton without or with missing energy.
This new feature of the UED model may be difficult to extract from the usual
 multijet plus multilepton signal of the SM, because of the huge
 SM background. Separation of signal from backgrounds is nontrivial.
 
On the other hand, hadronically quiet multilepton signals (may or may not be
accompanied by missing energy) are more interesting, though the initial 
production cross-section is suppressed by the ratio $\alpha/\alpha_s$.  
This signal may come from
the electroweak production processes noted earlier.
Let us look at the signal $n$ leptons plus missing energy, 
where $4\leq n \leq 7$.
For $R^{-1}=500$ GeV, the branching fractions are as follows:

\begin{table}[h]
\begin{center}
\begin{tabular}{|c|c|c|c|}
\hline
No    &       Processes           &                  final states            &     Branching  \\
         &                                  &                                                 &   (In percent)  \\
\hline
1      &       $W_1 W_1  $      &   6 l + $ {p_T}\slash$            &    15.1            \\
        &                                   &   4 l + ${p_T}\slash$                         &      3.0             \\
\hline
2      &       $Z_1 Z_1  $        &   6 l + ${p_T}\slash$                &      9.0               \\
        &                                  &   4 l + ${p_T}\slash$                   &          12.4                \\
%     &                    &   8 l                                    &                \\
\hline
3      &       $Z_1 B_1  $        &   4 l +  ${p_T}\slash$        &          21.2      \\
 %     &                    &   4 l + ${p_T}\slash$                   &                \\
\hline
4      &       $W_1 B_1  $       &   5 l + ${p_T}\slash$                   &     27.1           \\
\hline
5      &       $W_1 Z_1  $        &   5 l + ${p_T}\slash$                   &    21.2            \\
        &                                   &   7 l + ${p_T}\slash$                   &              15.1  \\
\hline
6      &       $l_1 L_1  $          &   4 l + ${p_T}\slash$                   &           6.1     \\
\hline
7      & ${l_1}$  ${L_1} $  &   4 l + ${p_T}\slash$                   &     6.1           \\
\hline
8      &       $n_1 N_1  $         &   4 l + ${p_T}\slash$                   &          15.1      \\
\hline
\end{tabular}
\caption{Branching fractions for multilepton final states.}
\end{center}
\end{table}

\begin{figure}[htbp]
%\centerline{
\hspace*{-0.2cm}
%\vskip -10pt
\rotatebox{-90}{\epsfxsize= 6.0 cm\epsfysize=8.0cm \epsfbox{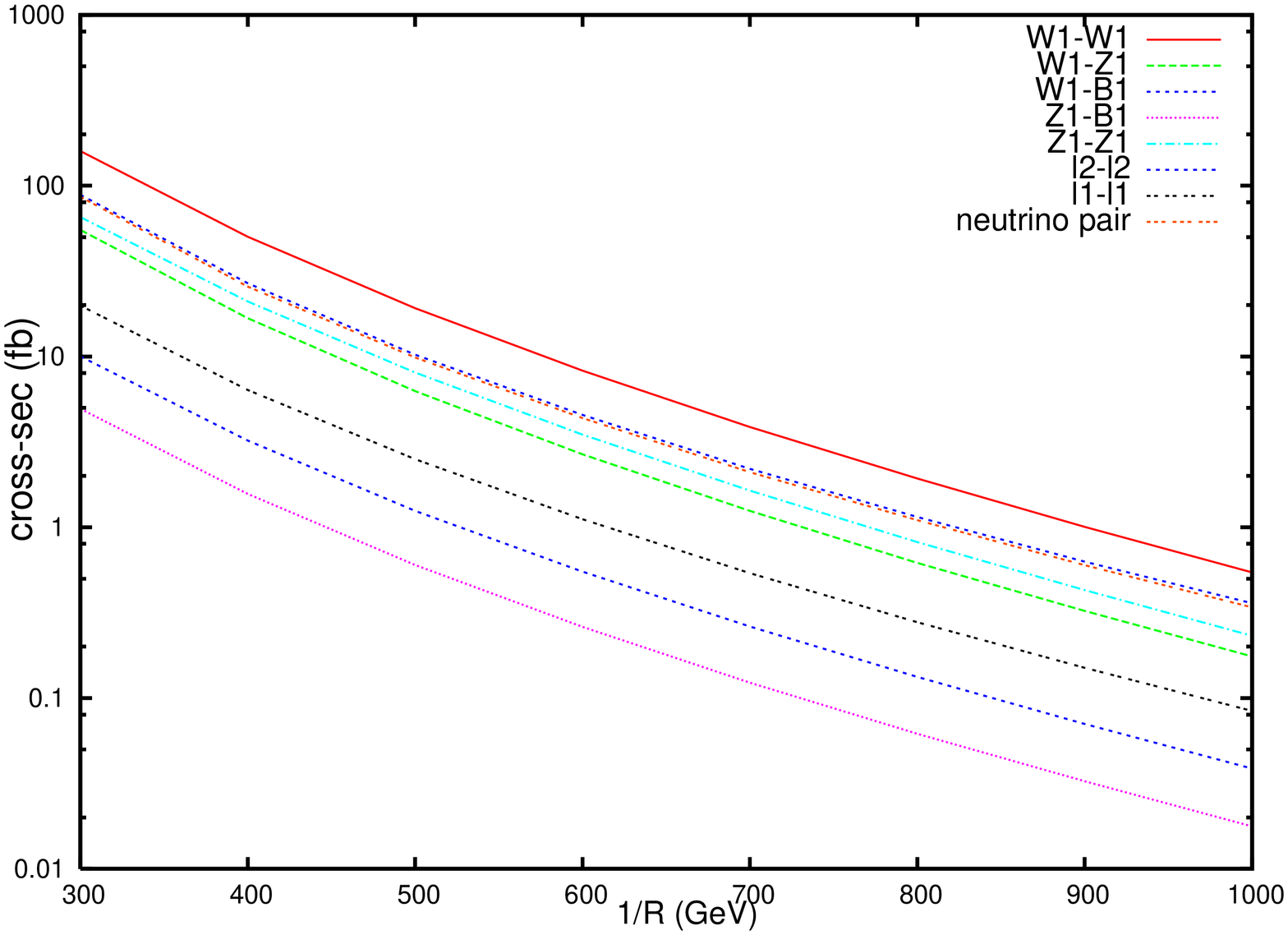}}
\hspace*{-0.2cm}
\rotatebox{-90}{ \epsfxsize= 6.0 cm\epsfysize=9.0cm \epsfbox{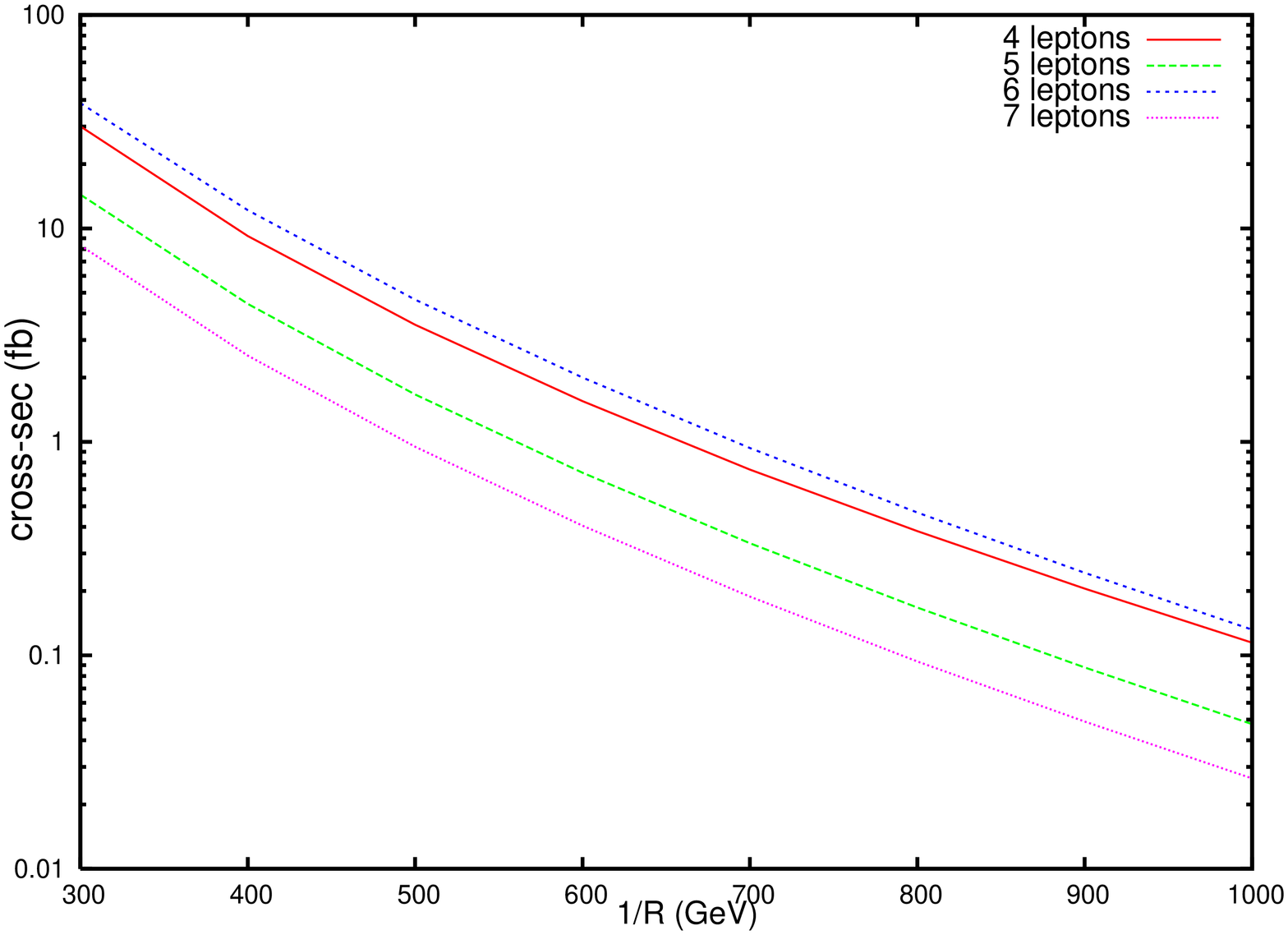} }
%}
\vskip -10pt
\caption{Left panel: cross-section for electroweak production at the LHC. 
Right panel: cross-section for multilepton plus missing energy signal.}
\label{fig:cross}
\end{figure}

We plot the variation of 4-7 leptons + missing energy cross section as a
function of $R^{-1}$ ($h$ does not play any role here except forcing the
decay of the LKP). The numerical cumputations were done with the
CalcHEP package \cite{Pukhov}, augmented by the implementation of UED.
The branching fractions, multiplied by the respective production cross-sections,
give the final signal cross-section and hence the event rate. 
It can be seen from figure \ref{fig:cross} that the $6\ell + p_T\slash$ signal
has the highest cross-section, closely followed by that of $4\ell+p_T\slash$.

The SM backgrounds are under comparative control. First, same-flavor unlike-sign
dilepton invariant mass veto at $M_Z$ removes the most important background
($WWZ \to 4\ell+p_T\slash$ is about 86 fb at the LHC). 
While the $W$-backgrounds, associated with neutrinos, cannot be removed in this
way, they are further suppressed by higher powers of $\alpha$. What may be 
problematic is to detect all the leptons coming from the excited states. Some of
them can be very soft, coming from decay between closely spaced levels, which
will probably missed by the acceptance of the detector. However, they come
from the KK-conserving decays at the first stage of the cascade. The KKPV decays
produce hard leptons, which should be easily detectable.  
In short, one should be able to discriminate such a scenario from other
competing new physics scenarios, as well as from the SM itself, from the event
rate and topology of the multilepton final state. A detailed study is outside 
the scope of this paper and will be taken up later.

\section{Summary}

In this paper we have discussed the phenomenology of KK parity violation in
the UED model through an asymmetrically fixed-point located term. This term
removes the $\ztwo$ parity of $y\to y+\pi R$ and hence KK parity is violated. 
Thus, the lightest $n=1$ particle, the LKP, is no longer stable and a cold dark
matter candidate. This removes the overclosure bound and answers the question 
posed in the Introduction, {\em viz.}, whether non-observation of UED signals
at the LHC will invalidate this model, in the negative. 
This also allows the introduction of gravitons in the model even for
comparatively low values of $R^{-1}$. 

The model that we have discussed is rather simplistic. The KKPV strength may
be different for different flavors, and that may give rise to interesting 
flavor-changing constraints, but we have assumed the same $h$ for all fermions. 
We have also kept the coupling small so that the field expansion is valid and
there is no single production of the excited states at the colliders. 

The removal of dark matter LKP opens up various possibilities in the LKP-NLKP
phase diagram, which we have studied. This, in turn, corresponds to different 
types of signals in the colliders.  
Depending on the strength of the coupling the LKP may decay inside or outside
the detector. If it decays inside the detector, the multilepton final states
should be useful to prove the validity of this model. 
For large values of the
SM Higgs boson mass and a small KK-parity violating coupling $h\sim 10^{-8}$,
$H^\pm_1$, which becomes a long-lived LKP to decay outside the detector, 
leaves its characteristic charged track.

\centerline{\bf {Acknowledgements}}
The author thanks Anirban Kundu for many useful discussions. He is
supported by a research fellowship of UGC, Govt.\ of India, and also
acknowledges the Regional Centre for Accelerator-based Particle Physics 
(ReCAPP), HRI, Allahabad, for a young associateship.

\end{document}